\documentclass [6pt,a4paper]{article}
\usepackage{bbm}
\usepackage{amsfonts}
\usepackage{mathrsfs}
\usepackage {amssymb}
\usepackage{cite}
\usepackage {amsmath}
\usepackage{color}
\usepackage{amsthm}
\usepackage{latexsym}

\usepackage{lastpage}
\usepackage{epsfig}

\begin{document}
	\newtheorem{lemma}{Lemma}[section]
	\newtheorem{theorem}[lemma]{Theorem}
	\newtheorem{example}[lemma]{Example}
	\newtheorem{definition}[lemma]{Definition}
	\newtheorem{proposition}[lemma]{Proposition}
	\newtheorem{corollary}[lemma]{Corollary}
	\newtheorem{case}[lemma]{Case}
	\newtheorem{remark}{Remark}
\begin{center}
\textbf{\large{On  cyclic codes of length $2^e$ over finite fields}}\footnote { E-mail
addresses:
 pbbmath@126.com(B.Pang),  zhushixin@hfut.edu.cn(S.Zhu), lpmath@126.com(P.Li). This research was supported by the National Natural Science Foundation of China under Grant No 61772168, 61572168 and 11501156.}\\
\end{center}

\begin{center}
{  Binbin Pang, Shixin Zhu, Ping Li}
\end{center}

\begin{center}
\textit{School of Mathematics, Hefei University of
Technology, Hefei 230009, Anhui, P.R.China }
\end{center}

\noindent\textbf{Abstract:} Professor Cunsheng Ding gave cyclotomic constructions of cyclic codes with length being the product of two primes. In this paper,  we study the cyclic codes of length $n=2^e$ and dimension $k=2^{e-1}$.  We  describe two new types of generalized cyclotomy of order two, which are different from Ding's. Furthermore, we  study two classes of cyclic codes of length $n$ and dimension $k$. We get the enumeration of these cyclic codes.  What's more, all of the codes from our construction are among the best cyclic codes. Furthermore, we study the hull of cyclic codes of length $n$ over $\mathbb{F}_q$. We obtain the range of $\ell=\dim({\rm Hull}(C))$. We construct and enumerate cyclic codes of length $n$ having hull of given dimension. We obtain some optimal cyclic codes with small dimensional hull       

\section{Introduction}
\hspace{1.2 em}Throughout this paper, let $\mathbb{F}_q$ be a finite field, where $q$ is an odd prime power.  Hamming weight of codeword $\mathbf{c}=(c_0, c_1, \cdots c_{n-1})\in C$ is the number of nonzero $c_i$ , and denoted by wt$(\mathbf{c})$. If $C$ is a $k$-dimensional subspace of $\mathbb{F}_q^n$, then $C$ is referred as an $[n,k,d]$ linear code over $\mathbb{F}_q$. If $d=n-k+1$, the code $C$ is called MDS code. The (Euclidean) dual code of $C$, denoted by $C^\perp$, is defined by $$C^\perp=\{ \mathbf{u}\in\mathbb{F}_q^n~|~\mathbf{u}\cdot \mathbf{c}=0~\forall~\mathbf{c}\in C\},$$ where  $\mathbf{u}\cdot \mathbf{c}$ denotes the standard inner product. The hull of linear code $C$ is defined by $${\rm Hull}(C):=C\cap C^\perp.$$

If a linear code $C$ is invariant under cyclic shift  on $\mathbb{F}_q$, $(c_0,c_1,\cdots, c_{n-1})\mapsto ( c_{n-1}, c_0,\cdots, c_{n-2}),$ then  $C$ is called a cyclic code.  By identifying a codeword with its polynomial representation in $\mathbb{F}_q[x]/\langle x^n-1\rangle$, a linear code over $\mathbb{F}_q$ of length $n$ is cyclic code if and only if the corresponding set in $\mathbb{F}_q[x]/\langle x^n-1\rangle$ is just an ideal in $\mathbb{F}_q[x]/\langle x^n-1\rangle$. Let $C$ be a cyclic code over $\mathbb{F}_q$. Then there is a monic polynomial $g(x)$ of minimal degree in $C$ such that $C=\langle g(x)\rangle$, where $g(x)|~(x^n-1)$. Furthermore, dim$(C)=n-$deg$g(x)$.

Since cyclic codes have efficient encoding and decoding algorithms, they have wide application in storage and communication systems. The cyclic codes were investigated in many literatures \cite{ref8,ref9,ref10,ref11,ref12}. But the enumeration of cyclic codes is in general unknown. 

Classical cyclotomy was considered by Gauss in his Disquisitiones Arithmeticae \cite{ref13} and cyclotomy has a wide application in literatures \cite{ref3,ref4,ref14,ref15,ref7}. The cyclotomic cyclic code is a simple construction of the best cyclic code \cite{ref1,ref2,ref5,ref6}. Ding and Pless  presented a cyclotomic approach to the construction of all binary duadic codes of prime lengths \cite{ref1}. Ding and Helleseth gave the new generalized cyclotomy \cite{ref16} and then they studied generalized cyclotomic codes of length $p_1^{e_1}\cdots p_t^{e_t}$  and the bound on the minimum odd-like weight of these codes with length $n=p_1^{e_1} p_2^{e_2}$ \cite{ref2}. After then, Ding gave three types of generalized cyclotomy and studied three classes of cyclic codes of length $n$ being the product of two primes and dimension $(n+1)/2$ and proved bound on their minimum odd-like weight\cite{ref5}. Recently, Xiong explained some of the numerical data by developing a general method on cyclic codes of composite length and on estimating the minimum distance \cite{ref6}.

The hull has been studied in many literatures \cite{ref18,ref19,ref20,ref21,ref22,ref23,ref24,ref25,ref26}. It is well known that the hull is important in many fields, especially when the hull is small \cite{ref18,ref19,ref20,ref21}. Sangwisut et. al. obtained the enumerations of cyclic and negacyclic codes of length $n$ over $\mathbb{F}_q$ having hulls of a given dimension \cite{ref26}. Recently, Li and Zeng presented sufficient and necessary condition that linear codes and cyclic codes have one-dimensional hull, and they constructed some linear codes and cyclic codes with one-dimension hull \cite{ref24}.  Inspired by these latter works, we consider the cyclic codes of length $n$ over finite field $\mathbb{F}_q$.
 
In this paper, we give some background and recall some basic results in Section 2. In Sections
3 and 4, we get the enumeration of cyclic codes $C_{(n,\frac{n}{2},q)}$ over $\mathbb{F}_q$.  We also  give a new cyclotomy of order two and construct some best cyclic codes with $q=2^sf+1$ and $q=2^sf+3$, respectively. In Section 5, we   construct and enumerate cyclic codes of length $n$ having given dimensional hull.  We conclude the paper in Section 6.

We will compare some of the codes presented in this paper with the tables of best known linear codes
(referred to as the Databse later) maintained by Markus Grassl at http://www.codetab les.de.  The example in this paper  is computed by Magma.

\section{Preliminaries}
\hspace{1.2 em}  In this paper, we always assume that gcd$(n,q)=1$. A cyclic code of length $n$ over $\mathbb{F}_q$ is referred as a simple-root cyclic code if $n$ and $q$ are relatively coprime, otherwise a repeated-root cyclic code. Let $\xi$ be a primitive $n$-th root of unity. Then we have 
$$x^n-1=\prod_{i\in\mathbb{Z}_{n}}(x-\xi^i).$$

Let $\mathbb{Z}_{n}$ denote the ring  $\{0, 1,\cdots,n-1\}$ with integer addition and integer multiplication modulo $n$.  For each $a\in\mathbb{Z}_{n}$, denote by $C_a$ the $q$-cyclotomic coset of $q$ modulo $n$ containing $a$. $C_a=\{a,aq,aq^2,\cdots,aq^{l_a-1}\}$ mod $n$, where $l_a$ is the smallest positive integer such that $a \equiv aq^{l_a}$ (mod $n$), and it is the size of the $q$-cyclotomic coset.  The smallest integer in $C_a$ is called the coset leader of $C_a$.  Let $\Gamma_{(q,n)}$ be the set of all the coset leaders. Then we have $$\mathbb{Z}_{n}=\bigcup_{a\in\Gamma_{(q,n)}}C_a.$$ 

It is well know that $m_{a}(x)=\Pi_{i\in C_a}(x-\xi^i)$ is a monic irreducible polynomial over $\mathbb{F}_q$, which is called minimum polynomial of $\xi^a$ over $\mathbb{F}_q$. Then we have $$x^n-1=\prod_{a\in\Gamma_{(q,n)}}m_a(x),$$ which is the factorization of $x^n-1$ into irreducible factors
over $\mathbb{F}_q$. This canonical factorization of  $x^n-1$ over $\mathbb{F}_q$
is crucial for the study of cyclic codes. Then we only need to choose the cyclotomic cosets to obtain the generator polynomial of cyclic codes $g(x)$. The set $S=~\{0\leq i\leq n-1~|~g(\xi^i)=0\}$ is referred to as the defining set of $C$.

It is well known that $\mathbb{Z}_{n}$ is the non-negative minimal complete set of residues modulo $n$. Let $\mathbb{Z}_{n}^\ast$ denote all the invertible elements of $\mathbb{Z}_{n}$. Denoted $\mathbb{Z}_{\bar{n}}$ by a reduced set of residues modulo $n$. Then we have  $$\mathbb{Z}_{\bar{n}}=\{z\in\mathbb{Z}_{n}|~{\rm gcd}(z,n)=1\}.$$

We recall the definition of the Euler function $\varphi$. For $n\in\mathbb{Z}^+$, $\varphi(n)$ is defined to be the number of integers between $1$ and $n$ relatively prime to $n$. Then we  give following lemmas, which can be found in literature \cite{ref17}. 
\begin{lemma}\label{le:2.1}
	Let symbols be the same as before. Then $$\mathbb{Z}_{n}^\ast=\mathbb{Z}_{\bar{n}}.$$ What's more
	$$\big|\mathbb{Z}_{n}^\ast\big|=\varphi(n),$$ where $\big|\mathbb{Z}_{n}^\ast\big|$ denotes the cardinality of $\mathbb{Z}_{n}^\ast$.
\end{lemma}

\begin{lemma}\label{le:2.2}
	Let $m>1$, then $m$ has primitive root if and only if $m$ is one of following from
	$$2,~4,~p^l,~2p^l,$$
	where $l\geq1$ and $p$ is an odd prime.
\end{lemma}

For any sets $A$ and $B$, define $$A\backslash B=\{x\in A~|~x\notin B\}.$$ 

\section{Cyclic codes $C_{(n,\frac{n}{2},q)}$ over $\mathbb{F}_q$ with $q=2^sf+1$}
\hspace{1.2 em} Throughout this section let $n=2^e$, $q=2^sf+1$ and $t=e-s$, where $e\geq3$, $s\geq2$ and $f$ is an odd integer. In this section, let $C_{(n,\frac{n}{2},q)}$ denote the code with length $n$ and dimension $\frac{n}{2}$ over $\mathbb{F}_q$. 
\subsection{All cyclic codes $C_{(n,\frac{n}{2},q)}$ over $\mathbb{F}_q$}
\hspace{1.2 em} In this subsection, we study all cyclic codes $C_{(n,\frac{n}{2},q)}$ by the generator polynomial. Assume that the multiplicative order of $q$ modulo $n$ is $r$, i.e., $r=$ord$_{n}(q)=2^{t}$. In the previous section, we give the definitions of $\mathbb{Z}_{n}$ and $\mathbb{Z}_{n}^\ast$. Next we give  definition of new sets as follows $$T_i=\{~2^ik|~1\leq k\leq 2^{e-i}~~ {\rm and} ~k~ {\rm is~ odd}~\}, ~{\rm for}~ 0\leq i\leq e-1,$$ let $T_e=\{0\}$ and we have $\big|T_i\big|=2^{e-i-1}$ ($0\leq i\leq e-1$). Then we have the following lemma immediately.

\begin{lemma}\label{le:3.1}
	Let symbols be the same as before. Then $$\mathbb{Z}_{n}=\bigcup_{i=0}^{e}T_i.$$
\end{lemma}
\proof For any $a\in\bigcup_{i=0}^{e}T_i$, we have $a\in\mathbb{Z}_{n}$. Then we have $\bigcup_{i=0}^{e}T_i\subseteq\mathbb{Z}_{n}$. Note that $T_i\bigcap T_j=\emptyset$ for $i\neq j$, $0\leq i,j\leq e$. Furthermore, $\sum_{i=0}^{e}\big|T_i\big|=n$. Then $\mathbb{Z}_{n}=\bigcup_{i=0}^{e}T_i.$\qed

From the definition of $T_i$, we give following lemma, which will be used in the sequel.
\begin{lemma}\label{le:3.2}
	For any $a\in T_i$, then we have
	
	(1) If $0\leq i\leq t-1$, $C_a\subseteq T_i~{\rm and}~\big|C_a\big|=2^{t-i}$;
	
	(2) If $t\leq i\leq e$, $C_a\subseteq T_i~{\rm and}~\big|C_a\big|=1$.
\end{lemma}  
\proof (1) For any $a=2^ik_0\in T_i$, where $1\leq k_0\leq 2^{e-i}$  and $k_0$ is odd. Then we obtain $qa\in T_i$, Thus $C_a\subseteq T_i$. Clearly, we have $\big|C_a\big|=2^{t-i}.$

(2) The previous subsection is similar to (1). Note that  $\big|C_a\big|=1.$  \qed

\begin{theorem}\label{th:3.3}
	Let symbols be the same as before. Then the total number of cyclic codes with length $n$ over $\mathbb{F}_q$ is equal to $$2^{t2^{s-1}+2^s}.$$
\end{theorem}
\proof From Lemmas $\ref{le:3.1}$ and $\ref{le:3.2}$, we obtain the total number of cyclotomic cosets is $$t2^{s-1}+2^s.$$ Thus we can easily get the desired conclusion.\qed

We always assume that $\binom{n}{0}=1$. Then we obtain the following theorem.
\begin{theorem}\label{th:3.4}
	Let symbols be the same as before. Then the total number of cyclic codes with length $n$ and dimension $\frac{n}{2}$ over $\mathbb{F}_q$ is equal to
	\begin{equation*} \sum_{\sum_{i=0}^{t}j_i2^{t-i}=2^{e-1}}\binom{2^{s-1}}{j_0}\binom{2^{s-1}}{j_1}\cdots\binom{2^{s-1}}{j_{t-1}}\binom{2^{s}}{j_t},\tag{1}
	\end{equation*} 
	where  $0\leq j_i\leq 2^{s-1}$, $0\leq i \leq t-1$ and $0\leq j_t\leq 2^{s}$.
\end{theorem}
\proof From Lemmas $\ref{le:3.1}$ and $\ref{le:3.2}$, we have the number of cyclotomic cosets with $\big|C_a\big|=2^{t-i}$ is $2^{s-1}$, for $0\leq i\leq t-1$, and the number of cyclotomic cosets with $\big|C_a\big|=1$ is $2^s$. Next we choose $j_i$ cyclotomic cosets with $\big|C_a\big|=2^{t-i}$ such that $\sum_{i=0}^{t}j_i2^{t-i}=2^{e-1}$, for $0\leq i\leq t$.  Thus we can easily get the desired conclusion.\qed

In order to show that all of the codes from our construction  are the best cyclic codes, we provide
information about all cyclic codes of length 8 and dimension 4 over $\mathbb{F}_5$, and all codes of
length 16 and dimension 8  over $\mathbb{F}_5$ in the sequel.\\
\\
A.  All cyclic codes $C_{(8,4,5)}$

From formula (1), we know that the total number of $C_{(8,4,5)}$ is 14. We obtain the factorization of $x^8-1$ over $\mathbb{F}_{5}$ as follows $$x^8-1=f_{11}(x)f_{12}(x)f_{13}(x)f_{14}(x)f_{21}(x)f_{22}(x),$$where polynomial $f_{ij}(x)$ as follows
\begin{equation*}
\begin{aligned} 
&f_{11}(x)=x+1 & &f_{12}(x)=x + 2&&f_{13}(x)=x + 3 \\ &f_{14}(x)=x + 4& &f_{21}(x)=x^2 + 2
& & f_{22}(x)=x^2 + 3\\
\end{aligned}
\end{equation*}

Let $g(x)$ and $d$ denote the generator polynomial and minimum distance of cyclic codes, respectively. Then we list all cyclic codes in Table 1.
\begin{table}[htbp]
	\center
	\caption{All cyclic codes  $C_{(8,4,5)}$}
	\begin{tabular}{cc|cc|cc|cc}
		\hline
		$g(x)$ & $d$  & $g(x)$ & $d$ & $g(x)$ & $d$ & $g(x)$ & $d$ \\
		\hline
		$f_{11}f_{12}f_{13}f_{14}$ & 2 &$f_{11}f_{12}f_{21}$ & 4 
		& $f_{11}f_{12}f_{22}$ & 4
		& $f_{11}f_{13}f_{21}$ &  4\\
		$f_{11}f_{13}f_{22}$ & 4 &$f_{11}f_{14}f_{21}$ & 3& $f_{11}f_{14}f_{22}$ &  3& $f_{12}f_{13}f_{21}$ &  3\\
		$f_{12}f_{13}f_{22}$ & 3 &$f_{12}f_{14}f_{21}$ &  4 & $f_{12}f_{14}f_{22}$ &  4& $f_{13}f_{14}f_{21}$ &  4\\
		$f_{13}f_{14}f_{22}$ & 4 &$f_{21}f_{22}$ & 2 & $-$ &  $-$&$-$ &  $-$\\
		\hline
	\end{tabular}
\end{table}\\
\\
B.  All cyclic codes $C_{(16,8,5)}$

From formula (1), we know that the total number of $C_{(16,8,5)}$ is 30. We obtain the factorization of $x^{16}-1$ over $\mathbb{F}_{5}$ as follows $$x^{16}-1=f_{11}(x)f_{12}(x)f_{13}(x)f_{14}(x)f_{21}(x)f_{22}(x)f_{41}(x)f_{42}(x),$$where polynomial $f_{ij}(x)$ as follows
\begin{equation*}
\begin{aligned} 
&f_{11}(x)=x+1 & &f_{12}(x)=x + 2&&f_{13}(x)=x + 3 & 
&f_{14}(x)=x + 4&\\ &f_{21}(x)=x^2 + 2
& & f_{22}(x)=x^2 + 3&
&f_{41}(x)=x^4 + 2& &f_{42}(x)=x^4 + 3
& & \\
\end{aligned}
\end{equation*}

Let $g(x)$ and $d$ denote the generator polynomial and minimum distance of cyclic codes, respectively. Then we list all cyclic codes in Table 2.
\begin{table}[htbp]
	\center
	\caption{All cyclic codes  $C_{(16,8,5)}$}
	\begin{tabular}{cc|cc|cc|cc}
		\hline
		$g(x)$ & $d$  & $g(x)$ & $d$ & $g(x)$ & $d$ & $g(x)$ & $d$ \\
		\hline
		$f_{11}f_{12}f_{13}f_{14}f_{21}f_{22}$ & 2 &$f_{11}f_{12}f_{13}f_{14}f_{41}$ & 3
		&$f_{11}f_{12}f_{13}f_{14}f_{42}$ & 3
		& $f_{11}f_{12}f_{21}f_{41}$ &  4\\
		$f_{11}f_{12}f_{21}f_{42}$ & 4 &$f_{11}f_{12}f_{22}f_{41}$ & 4& $f_{11}f_{12}f_{22}f_{42}$ &  4& $f_{11}f_{13}f_{21}f_{41}$ &  4\\
		$f_{11}f_{13}f_{21}f_{42}$ & 4 &$f_{11}f_{13}f_{22}f_{41}$ & 4 
		&$f_{11}f_{13}f_{22}f_{42}$ & 4
		& $f_{11}f_{14}f_{21}f_{41}$ &  4\\
		$f_{11}f_{14}f_{21}f_{42}$ & 4 &$f_{11}f_{14}f_{22}f_{41}$ & 4& $f_{11}f_{14}f_{22}f_{42}$ &  4& $f_{12}f_{13}f_{21}f_{41}$ &  4\\
		$f_{12}f_{13}f_{21}f_{42}$ & 4 &$f_{12}f_{13}f_{22}f_{41}$ & 4 
		&$f_{12}f_{13}f_{22}f_{42}$ & 4
		& $f_{12}f_{14}f_{21}f_{41}$ &  4\\
		$f_{12}f_{14}f_{21}f_{42}$ & 4 &$f_{12}f_{14}f_{22}f_{41}$ & 4& $f_{12}f_{14}f_{22}f_{42}$ &  4& 
		$f_{13}f_{14}f_{21}f_{41}$ & 4\\ $f_{13}f_{14}f_{21}f_{42}$ & 4& $f_{13}f_{14}f_{22}f_{41}$ &  4& $f_{13}f_{14}f_{22}f_{42}$ & 4& 
		$f_{21}f_{22}f_{41}$ & 3\\
		$f_{21}f_{22}f_{42}$ & 3& $f_{41}f_{42}$ &  2& $-$ & $-$& $-$& $-$ \\
		\hline
	\end{tabular}
\end{table}

\subsection{A cyclotomy of order two and its codes}
\hspace{1.2 em} From Lemma \ref{le:2.2}, we know that $n=2^e$ doesn't have primitive root, where $e\geq3$. Then the Ding's construction in \cite{ref5} can not be used in this place. In this subsection, we give a new generalized cyclotomic classes $W_j$ of order $2$ by the number theory.  First of all, we give a important lemma from the number theory \cite{ref17}.

\begin{lemma} \label{le:4.1}
	Let $e\geq3$, $n=2^e$, then the following sentences are hold
	
	1) the degree of $5$ modulo $n$ is $2^{e-2}$;
	
	2) the reduced set of residues modulo  $n$ as follows $$\mathbb{Z}_{\bar{n}}=\big\{\pm5^b~({\rm mod}\ n)~|~0\leq b < 2^{e-2}\big\}.$$
\end{lemma}

Note that $\mathbb{Z}_{n}^\ast=\mathbb{Z}_{\bar{n}}=T_0$. Then we the definition of a generalized cyclotomic classes $W_i$ of order 2 as follows
$$W_i=\big\{(-1)^i5^b~|~0\leq b < 2^{e-2}\big\},~i=0,1.$$

Notice that $W_0$ is a subgroup of $\mathbb{Z}_{n}^\ast$ and $W_0=-W_1$. Thus the sets $W_0$ and $W_1$ form a cyclotomy of order 2. From Lemma \ref{le:3.1} and the proof of the Theorem \ref{th:3.3}.  Clearly, If $1\leq i\leq t-1$,  every set $T_i$ is a union of $2^{s-1}$ $q$-cyclotomic set $C_a$, where $\big|C_a\big|=2^{t-i}.$  If $t\leq i\leq e-1$, every set $T_i$ is a union of $2^{e-i-1}$ $q$-cyclotomic set $C_a$, where $\big|C_a\big|=1.$ In other words, there are $2^s$ $q$-cyclotomic sets $C_a$, where $\big|C_a\big|=1.$

Next we can use this cyclotomy ($W_0,W_1$) to construct some cyclic codes. We give a definition as follows
$$w_j(x)=\prod_{i\in W_j}(x-\xi^i),~j\in\{0,1\}.$$ 

\begin{lemma}\label{le:4.2}
	Let symbols be the same as before. Then  we have $w_j(x)\in \mathbb{F}_q[x]$, where $j\in\{0,1\}$.
\end{lemma}
\proof Since $q=2^sf+1$, $s\geq2$, we obtain $q\in W_0$. Then we have $qW_j = W_j$, where $j\in\{0,1\}$. It then follows that $$w_j(x)^q=\prod_{i\in W_j}(x^q-\xi^{qi})=\prod_{i\in qW_j}(x^q-\xi^{i})=w_j(x^q),$$then we have $w_j(x)\in \mathbb{F}_q[x]$.\qed 

Let $D_0^{(i)}$ is a union of any $2^{s-2}$ $q$-cyclotomic sets $C_a$, where $C_a\subseteq T_i$, and $D_1^{(i)}=T_i\setminus D_0^{(i)}$, for  $1\leq i\leq t-1$. Let $D_0^{(i)}$ is a union of any $2^{e-i-2}$ $q$-cyclotomic sets $C_a$, where $C_a\subseteq T_i$ and $\big|C_a\big|=1$, and $D_1^{(i)}=T_i\setminus D_0^{(i)}$ for $t\leq i\leq e-2$. Let $D_0^{(e-1)}=T_{e-1}$ and $D_1^{(e-1)}=T_e$. Next we define for each $1\leq i\leq e-1$,
\begin{equation*}
\begin{aligned} 
&d_0^{(i)}(x)=\prod_{j\in D_0^{(i)}}(x-\xi^j), \\ 
&d_1^{(i)}(x)=\prod_{j\in D_1^{(i)}}(x-\xi^j).\\
\end{aligned}
\end{equation*}

In fact, from Lemma \ref{le:3.1}, we have 
$$x^n-1=w_0(x)w_1(x)\prod_{ i=1} ^{e-1}d_0^{(i)}(x)d_1^{(i)}(x).$$ 

Next we construct cyclic codes from the cyclotomy $(W_0,W_1)$. We can obtain cyclic codes with generator polynomials as follows
\begin{equation*} 
g(x)=w_j(x)\prod_{ i=1} ^{e-1}d_{l_i}^{(i)}(x),\tag{2}
\end{equation*} 
where $j, l_i \in\{0,1\},$ $1\leq i\leq e-1.$

Obviously, the cyclic codes with generator polynomial $g(x)$ have length $n$ and dimension $\frac{n}{2}$. The total number of this cyclic codes with generator polynomial as $(2)$ is equal to 
\begin{displaymath} 
4\binom{2^{s-1}}{2^{s-2}}^{t-1}\prod_{i=t}^{e-2}\binom{2^{e-i-1}}{2^{e-i-2}}.
\end{displaymath}

From the construction and the definitions of $T_i$ and $g(x)$, the desired conclusion is obtained immediately. In the sequel, we give some example show that the cyclic codes from this construction are the best cyclic codes.
\begin{example}\label{e:1}
	Let $(e,s)=(3,2)$. Then from the construction we have 8 cyclic codes in Table 3.
\begin{table}[htbp]
	\center
	\caption{Best cyclic codes with  $(e,s)=(3,2)$}
	\begin{tabular}{cc|cc|cc|cc}
		\hline
		$g(x)$ & $d$  & $g(x)$ & $d$ & $g(x)$ & $d$ & $g(x)$ & $d$ \\
		\hline
		$f_{11}f_{12}f_{21}$ & 4 &$f_{11}f_{12}f_{22}$ & 4 
		&$f_{11}f_{13}f_{21}$ & 4
		& $f_{11}f_{13}f_{22}$ &  4\\
		$f_{12}f_{14}f_{21}$ & 4 &$f_{12}f_{14}f_{22}$ & 4& $f_{13}f_{14}f_{21}$ &  4& 
		$f_{13}f_{14}f_{22}$ & 4\\ 
		\hline
	\end{tabular}
\end{table}
\end{example}

By comparing Tables 1 and 3, we know that all of cyclic codes from our construction are the best cyclic codes.
\begin{example}
	Let $(e,s)=(4,2)$. Then from the construction we have 16 cyclic codes in Table 4.
	\begin{table}[htbp]
		\center
		\caption{Best cyclic codes with $(e,s)=(4,2)$}
		\begin{tabular}{cc|cc|cc|cc}
			\hline
			$g(x)$ & $d$  & $g(x)$ & $d$ & $g(x)$ & $d$ & $g(x)$ & $d$ \\
			\hline
			$f_{11}f_{12}f_{21}f_{41}$ & 4 &$f_{11}f_{12}f_{21}f_{42}$ & 4 
			&$f_{11}f_{12}f_{22}f_{41}$ & 4
			& $f_{11}f_{12}f_{22}f_{42}$ &  4\\
			$f_{11}f_{13}f_{21}f_{41}$ & 4 &$f_{11}f_{13}f_{21}f_{42}$ & 4& $f_{11}f_{13}f_{22}f_{41}$ &  4& 
			$f_{11}f_{13}f_{22}f_{42}$ & 4\\
            $f_{12}f_{14}f_{21}f_{41}$ & 4 &$f_{12}f_{14}f_{21}f_{42}$ & 4 
            &$f_{12}f_{14}f_{22}f_{41}$ & 4
            & $f_{12}f_{14}f_{22}f_{42}$ &  4\\
            $f_{13}f_{14}f_{21}f_{41}$ & 4 &$f_{13}f_{14}f_{21}f_{42}$ & 4& $f_{13}f_{14}f_{22}f_{41}$ &  4& 
            $f_{13}f_{14}f_{22}f_{42}$ & 4\\  
			\hline
		\end{tabular}
	\end{table}
\end{example}
By comparing Tables 2 and  4, we know that all of cyclic codes from our construction are the best cyclic codes.
\section{Cyclic codes $C_{(n,\frac{n}{2},q)}$ over $\mathbb{F}_q$ with $q=2^sf+3$}
\hspace{1.2 em} Throughout  this section let $n=2^e$, $q=2^sf+3$, where $e\geq3$, $s\geq3$ and $f$ is an odd integer. In this section, let $C_{(n,\frac{n}{2},q)}$ denote the code with length $n$ and dimension $\frac{n}{2}$ over $\mathbb{F}_q$.
\subsection{All cyclic codes $C_{(n,\frac{n}{2},q)}$ over $\mathbb{F}_q$}
\hspace{1.2 em}In this subsection, we study all cyclic codes $C_{(n,\frac{n}{2},q)}$ by the generator polynomial.  In the previous section, we give the definitions of $T_i$, $\mathbb{Z}_{n}$ and $\mathbb{Z}_{n}^\ast$. Then we have the following lemma, which will be used in the sequel.

\begin{lemma}\label{le:5.1}
	For any $a\in T_i$, then we have $C_a\subseteq T_i$, what's more
	
	(1) If $0\leq i\leq e-3$, $\big|C_a\big|=2^{e-i-2}$;
	
	(2) $\big|C_{e-2}\big|=2$ and $\big|C_{e-1}\big|=1$.
\end{lemma}  
\proof For any $a=2^ik_0\in T_i$, where $1\leq k_0\leq 2^{e-i}$  and $k_0$ is odd. Then we obtain $qa\in T_i$, Thus $C_a\subseteq T_i$. Clearly, we have $\big|C_a\big|=2^{e-i-2}.$ Note that  $\big|C_{e-2}\big|=2$ and $\big|C_{e-1}\big|=1$.  \qed

\begin{theorem}\label{th:5.2}
	Let symbols be the same as before. Then the total number of cyclic codes with length $n$ over $\mathbb{F}_q$ is equal to $$2^{2e-1}.$$
\end{theorem}
\proof From Lemmas $\ref{le:3.1}$ and $\ref{le:5.1}$, we obtain the total number of cyclotomic cosets is $$2e-1.$$ Thus we can easily get the desired conclusion.\qed

Then we obtain the following theorem.
\begin{theorem}\label{th:5.3}
	Let symbols be the same as before. Then the total number of cyclic codes with length $n$ and dimension $\frac{n}{2}$ over $\mathbb{F}_q$ is equal to
	\begin{equation*} \sum_{\sum_{i=0}^{e-2}j_i2^{e-i-2}=2^{e-1}}\binom{2}{j_0}\binom{2}{j_1}\cdots\binom{3}{j_{e-3}}\binom{2}{j_{e-2}},\tag{3}
	\end{equation*} 
	where  $0\leq j_i\leq 2$ when $0\leq i \leq e-4$ or $i=e-2$ and $0\leq j_{e-3}\leq 3$.
\end{theorem}
\proof From Lemmas $\ref{le:3.1}$ and $\ref{le:5.1}$, we have the number of cyclotomic cosets with $\big|C_a\big|=2^{e-i-2}$ is $2$, for $0\leq i\leq e-4$, the number of cyclotomic cosets with $\big|C_a\big|=2$ is $3$ and the number of cyclotomic cosets with $\big|C_a\big|=1$ is $2$. Next we choose $j_i$ cyclotomic cosets with $\big|C_a\big|=2^{e-i-2}$ such that $\sum_{i=0}^{e-2}j_i2^{e-i-2}=2^{e-1}$.  Thus we can easily get the desired conclusion.\qed

Next we give some examples as follows.\\
\\
A.  All cyclic codes $C_{(16,8,11)}$

From formula (3), we know that the total number of $C_{(16,8,11)}$ is 14. We obtain the factorization of $x^{16}-1$ over $\mathbb{F}_{11}$ as follows $$x^{16}-1=f_{11}(x)f_{12}(x)f_{21}(x)f_{22}(x)f_{23}(x)f_{41}(x)f_{41}(x),$$where polynomial $f_{ij}(x)$ as follows
\begin{equation*}
\begin{aligned} 
&f_{11}(x)=x+1 &&f_{12}(x)=x + 10& &f_{21}(x)=x^2 + 1&&f_{22}(x)=x^2 + 3x+10 \\ &f_{23}(x)=x^2 +8x+ 10&&f_{41}(x)=x^4 + 3x^2+10
&& f_{42}(x)=x^4+8x^2 + 10\\
\end{aligned}
\end{equation*}

Let $g(x)$ and $d$ denote the generator polynomial and minimum distance of cyclic codes, respectively. Then we list all cyclic codes in Table 5.
\begin{table}[htbp]
	\center
	\caption{All cyclic codes $C_{(16,8,11)}$}
	\begin{tabular}{cc|cc|cc|cc}
		\hline
		$g(x)$ & $d$  & $g(x)$ & $d$ & $g(x)$ & $d$ & $g(x)$ & $d$ \\
		\hline
		$f_{11}f_{12}f_{21}f_{22}f_{23}$ & 2 &$f_{11}f_{12}f_{21}f_{41}$ & 4 
		& $f_{11}f_{12}f_{21}f_{42}$ & 4
		& $f_{11}f_{12}f_{22}f_{41}$ &  6\\
		$f_{11}f_{12}f_{22}f_{42}$& 6 &$f_{11}f_{12}f_{23}f_{41}$ & 6&	$f_{11}f_{12}f_{23}f_{42}$ &  6& $f_{21}f_{22}f_{41}$ &  6\\
		$f_{21}f_{22}f_{42}$ & 6 &$f_{21}f_{23}f_{41}$ &  6 & $f_{21}f_{23}f_{42}$ &  6& $f_{22}f_{23}f_{41}$ &  4\\
		$f_{22}f_{23}f_{42}$ & 4 &$f_{41}f_{42}$ & 2 & $-$ &  $-$&$-$ &  $-$\\
		\hline
	\end{tabular}
\end{table}\\
\\
B.  All cyclic codes $C_{(32,16,19)}$

From formula (3), we know that the total number of $C_{(32,16,19)}$ is 30. We obtain the factorization of $x^{32}-1$ over $\mathbb{F}_{19}$ as follows $$x^{32}-1=f_{11}(x)f_{12}(x)f_{13}(x)f_{14}(x)f_{21}(x)f_{22}(x)f_{41}(x)f_{42}(x),$$where polynomial $f_{ij}(x)$ as follows
\begin{equation*}
\begin{aligned} 
&f_{11}(x)=x+1 & &f_{12}(x)=x + 18&&f_{21}(x)=x^2 + 1 \\ 
&f_{22}(x)=x^2 + 6x+18& &f_{23}(x)=x^2 +13x +18
& & f_{41}(x)=x^4 + 6x^2+18\\
&f_{42}(x)=x^4 + 13x^2+18& &f_{81}(x)=x^8 + 6x^4+18
& &f_{82}(x)=x^8 + 13x^4+18 \\
\end{aligned}
\end{equation*}

Let $g(x)$ and $d$ denote the generator polynomial and minimum distance of cyclic codes, respectively. Then we list all cyclic codes in Table 6.
\begin{table}[htbp]
	\center
	\caption{All cyclic codes $C_{(32,16,19)}$}
	\begin{tabular}{cc|cc|cc|cc}
		\hline
		$g(x)$ & $d$  & $g(x)$ & $d$ & $g(x)$ & $d$ & $g(x)$ & $d$ \\
		\hline
		$f_{81}f_{82}$ & 2 &$f_{11}f_{12}f_{21}f_{41}f_{81}$ & 6
		&$f_{11}f_{12}f_{21}f_{41}f_{82}$ & 6
		& $f_{11}f_{12}f_{21}f_{42}f_{81}$ &  6\\
		$f_{11}f_{12}f_{21}f_{42}f_{82}$ & 6 &$f_{11}f_{12}f_{22}f_{41}f_{81}$ & 6& $f_{11}f_{12}f_{22}f_{41}f_{82}$ &  6& $f_{11}f_{12}f_{22}f_{42}f_{81}$ &  6\\
	    $f_{11}f_{12}f_{22}f_{42}f_{82}$ & 6 &$f_{11}f_{12}f_{23}f_{41}f_{81}$ & 6 
		&$f_{11}f_{12}f_{23}f_{41}f_{82}$  & 6
		& $f_{11}f_{12}f_{23}f_{42}f_{81}$ &  6\\
		$f_{11}f_{12}f_{23}f_{42}f_{82}$ & 6 & $f_{11}f_{12}f_{21}f_{22}f_{23}f_{81}$ & 4& $f_{11}f_{12}f_{21}f_{22}f_{23}f_{82}$ &  4& $f_{21}f_{22}f_{41}f_{81}$ &  6\\
		$f_{21}f_{22}f_{41}f_{82}$ & 6 &$f_{21}f_{22}f_{42}f_{81}$ & 6 
		&$f_{21}f_{22}f_{42}f_{82}$ & 6
		& $f_{21}f_{23}f_{41}f_{81}$ &  6\\
		$f_{21}f_{23}f_{41}f_{82}$ & 6 &$f_{21}f_{23}f_{42}f_{81}$ & 6& $f_{21}f_{23}f_{42}f_{82}$ &  6& 
		$f_{22}f_{23}f_{41}f_{81}$ & 6\\ $f_{22}f_{23}f_{41}f_{82}$ & 6& 	$f_{22}f_{23}f_{42}f_{81}$ &  6& $f_{22}f_{23}f_{42}f_{82}$ & 6& 
		$f_{41}f_{42}f_{81}$ & 4\\
		$f_{41}f_{42}f_{82}$ & 4& $f_{11}f_{12}f_{21}f_{22}f_{23}f_{41}f_{42}$ &  2& $-$ & $-$& $-$& $-$ \\
		\hline
	\end{tabular}
\end{table}
\subsection{A new cyclotomy of order two and its codes}
\hspace{1.2 em} In this section, we give  a new generalized cyclotomic classes $U_k$ of order $2$ by the number theory as follows
$$U_0=\big\{5^i,~ (-1)5^j~|~i ~{\rm ~is~ even~ and}~ j~{\rm  is ~odd},~ 0\leq i, j < 2^{e-2} \big\},$$
$$U_1=\big\{5^i,~ (-1)5^j~|~i ~{\rm ~is~ odd~ and}~ j~{\rm  is ~even},~ 0\leq i, j < 2^{e-2} \big\}.$$

Notice that $U_0$ is a subgroup of $\mathbb{Z}_{n}^\ast$ and $U_0=-U_1$. Thus the sets $U_0$ and $U_1$ form a cyclotomy of order 2. From Lemma \ref{le:5.1}, we know that if  $1\leq i\leq e-3$,  every set $T_i$ is a union of $2$ $q$-cyclotomic set $C_a$.  If $e-2\leq i\leq e$, every set $T_i$ is a  $q$-cyclotomic set. 

Next we can use this cyclotomy ($U_0,U_1$) to construct some cyclic codes. We give a definition as follows
$$u_j(x)=\prod_{i\in U_j}(x-\xi^i),~j\in\{0,1\}.$$ 

\begin{lemma}\label{le:6.1}
	Let symbols be the same as before. Then  we have $u_j(x)\in \mathbb{F}_q[x]$, where $j\in\{0,1\}$.
\end{lemma}
\proof The proof is similar to Lemma \ref{le:4.2}. So we omit it here.  \qed

Let $D_0^{(i)}$ be a union of anyone $q$-cyclotomic set $C_a$, where $C_a\subseteq T_i$, and $D_1^{(i)}=T_i\setminus D_0^{(i)}$, for  $1\leq i\leq e-3$. Let $D^{(e-2)}=T_{e-2}$,  $D_0^{(e-1)}=T_{e-1}$ and $D_1^{(e-1)}=T_e$. Next we define for each $1\leq i\leq e-3$ and $i=e-1$,
\begin{equation*}
\begin{aligned} 
&d_0^{(i)}(x)=\prod_{j\in D_0^{(i)}}(x-\xi^j), \\ 
&d_1^{(i)}(x)=\prod_{j\in D_1^{(i)}}(x-\xi^j).\\
\end{aligned}
\end{equation*}
and $d^{(e-2)}(x)=\prod_{j\in D^{(e-2)}}(x-\xi^j).$
In fact, from Lemma \ref{le:3.1}, we have 
$$x^n-1=d^{(e-2)}(x)\prod_{j=0}^1u_j(x)d_j^{(e-1)}(x)\prod_{ i=1}^{e-3}d_0^{(i)}(x)d_1^{(i)}(x).$$ 

Next we construct cyclic codes from the cyclotomy $(U_0,U_1)$. We can obtain cyclic codes with generator polynomial as follows
\begin{equation*} 
g(x)=u_j(x)d^{(e-2)}(x)\prod_{ i=1}^{e-3}d_{l_i}^{(i)}(x)~
{\rm or}~u_j(x)\prod_{k=0}^1d_k^{(e-1)}(x)\prod_{ i=1}^{e-3}d_{l_i}^{(i)}(x)\tag{4}
\end{equation*} 
where $j, l_i \in\{0,1\},$ $0\leq i\leq e-3.$

Obviously, the cyclic codes with generator polynomial $g(x)$ have length $n$ and dimension $\frac{n}{2}$. The total number of this cyclic codes with generator polynomial as $(4)$ is equal to 
$$2^{e-1}.$$

From the construction and the definitions of $T_i$ and $g(x)$, the desired conclusion is obtained immediately. In the sequel, we give some examples show that the cyclic codes from this construction are the best cyclic codes.
\begin{example}
	Let $(e,s)=(4,3)$. Then from the construction we have 8 cyclic codes in Table 7.
	\begin{table}[htbp]
		\center
		\caption{Best cyclic codes with  $(e,s)=(4,3)$}
		\begin{tabular}{cc|cc|cc|cc}
			\hline
			$g(x)$ & $d$  & $g(x)$ & $d$ & $g(x)$ & $d$ & $g(x)$ & $d$ \\
			\hline
			$f_{11}f_{12}f_{22}f_{41}$ & 6 &$f_{11}f_{12}f_{22}f_{42}$ & 6 
			&$f_{11}f_{12}f_{23}f_{41}$ & 6
			& $f_{11}f_{12}f_{23}f_{42}$ &  6\\
	     	$f_{21}f_{22}f_{41}$ & 6 &$f_{21}f_{22}f_{42}$ & 6 
            &$f_{21}f_{23}f_{41}$ & 6
            & $f_{21}f_{23}f_{42}$ &  6\\
			\hline
		\end{tabular}
	\end{table}
\end{example}

By comparing Tables 5 and  7, we know that all of cyclic codes from our construction are the best cyclic codes.

\begin{example}
	Let $(e,s)=(5,4)$. Then from the construction we have 16 cyclic codes in Table 8.
	\begin{table}[htbp]
		\center
		\caption{Best cyclic codes with $(e,s)=(5,4)$}
		\begin{tabular}{cc|cc|cc|cc}
			\hline
			$g(x)$ & $d$  & $g(x)$ & $d$ & $g(x)$ & $d$ & $g(x)$ & $d$ \\
			\hline
$f_{11}f_{12}f_{22}f_{41}f_{81}$ & 6 &$f_{11}f_{12}f_{22}f_{41}f_{82}$ & 6 
&$f_{11}f_{12}f_{22}f_{42}f_{81}$ & 6
& $f_{11}f_{12}f_{22}f_{42}f_{82}$ &  6\\
$f_{11}f_{12}f_{23}f_{41}f_{81}$ & 6 &$f_{11}f_{12}f_{23}f_{41}f_{82}$ & 6 
&$f_{11}f_{12}f_{23}f_{42}f_{81}$ & 6
& $f_{11}f_{12}f_{23}f_{42}f_{82}$ &  6\\

$f_{21}f_{22}f_{41}f_{81}$ & 6 &$f_{21}f_{22}f_{41}f_{82}$ & 6 
&$f_{21}f_{22}f_{42}f_{81}$ & 6
& $f_{21}f_{22}f_{42}f_{82}$ &  6\\
$f_{21}f_{23}f_{41}f_{81}$ & 6 &$f_{21}f_{23}f_{41}f_{82}$ & 6 
&$f_{21}f_{23}f_{42}f_{81}$ & 6
& $f_{21}f_{23}f_{42}f_{82}$ &  6\\
			\hline
		\end{tabular}
	\end{table}
\end{example}
By comparing Tables 6 and 8, we know that all of cyclic codes from our construction are the best cyclic codes.
\section{The hull of cyclic codes with length $n$ over $\mathbb{F}_q$}
\hspace{1.2 em} In this section, we study the hull of cyclic codes with lengths $n$ by the defining set. We obtain all possible value of dimension of hull.  What's more, we construct some cyclic codes having given dimension and get the enumeration of this cyclic codes. Let $\ell$ denotes the dimension of Hull($C$).  

Let $S$ denotes the defining set of cyclic code $C$. We have the defining set of Hull($C$) is $S\bigcup(\mathbb{Z}_{n}\setminus(-S)),$ where $-S=\{n-x~|x\in S\}$.  From Theorem 4.1 in \cite{ref24}, we have following lemma immediately.

\begin{lemma}\label{le:9.1} Let $C$ be a cyclic code of length $n$ over $\mathbb{F}_q$ and let $S$ be its defining set, $\ell={\rm dim(Hull}($C$))$. Then the code $C$ has one-dimension hull if and only if there is a unique integer $i$ with $0\leq i\leq g-1$ such that $S\setminus(-S)=\{\frac{in}g\}$, where $g=\gcd(n,q-1)$. what's more, $\ell=|S\setminus(-S)|$.
\end{lemma}

\subsection{The hull of cyclic code when $q=2^sf+1$}
\hspace{1.2 em}In this subsection, let $q=2^sf+1$. Let symbols be the same as Section 3, we have the following lemma immediately.

\begin{lemma}\label{le:7.1} For any $a\in T_i$, $C_a\in T_i$. We have following sentences are hold

(1)  If $0\leq i\leq e-2$, $C_a \bigcap (-C_a)=\emptyset$;

(2)  If $e-1\leq i\leq e$, $C_a\bigcap (-C_a)=C_a$.
\end{lemma}
\proof (1) For $0\leq i\leq e-2$, if $C_a \bigcap (-C_a)\neq\emptyset$, then exist positive integer $j$ and $x=2^ik\in C_a$ such that $n-x\equiv xq^j~({\rm mod}~n)$, which is impossible. 

(2) For $e-1\leq i\leq e$, it is clearly that $C_a\bigcap (-C_a)=C_a$. \qed

In order to obtain the enumeration of  cyclic codes having given dimension. For any $C_a\in T_i$,  we give definition of new sets as follows

$$S_{ij}=\{C_a\bigcup-C_a~|~0\leq i\leq e-2, 1\leq j\leq2^{s-2}\},$$
and $S_{i0}=\{C_a~|~e-1\leq i\leq e\}$. Clearly, we have following lemma and omit the detail proof.
\begin{lemma}\label{le:7.2} For any $a\in T_i$, $C_a\in T_i$. We have following sentences are hold
	
	(1) $S_{ij}\bigcap(-S_{ij})=S_{ij}$;
	
	(2) $S_{ij}\bigcap(-S_{kl})=\emptyset$ if $(i,j)\neq(k,l)$;
	
	(3)  If $0\leq i\leq e-2$, $T_i=\bigcup_{j=1}^{2^{s-2}}S_{ij}$;
	
	(4)  If $e-1\leq i\leq e$, $T_i=S_{i0}$.
\end{lemma}
\begin{theorem}\label{th:7.3}
	Let symbols be the same as before. Then the total number of LCD cyclic codes with length $n$ over $\mathbb{F}_q$ is equal to
	$$2^{(t+2)2^{s-2}+1}.$$
\end{theorem}
\proof Since LCD cyclic code $C$ is equivalent to the dimension of Hull($C$) is zero, i.e., $\ell={\rm dim(Hull}(C))=0$.  Let the defining set of $C$ is $S$, then $\ell=|S\setminus(-S)|$. Then we have $S=-S$. Thus $S$ is a union of some set $S_{ij}$. From Lemma \ref{le:7.2}, we know that the total number of  $S_{ij}$ is $${(t+2)2^{s-2}+1},$$ this is complete the proof.\qed

\begin{theorem}\label{th:7.4}
Let symbols be the same as before. Let $C$ be cyclic codes with length $n$ over $\mathbb{F}_q$,  $S$ denote the defining set of cyclic code $C$ and $\ell={\rm dim(Hull}($C$)).$ Then we have 
	$$ \ell\in N_1=\{~\sum_{i=0}^{t}n_i2^i~|~0\leq n_0<2^{s-1}, 0\leq n_i\leq2^{s-2}~ for~ 1\leq i\leq t\},$$ where $n_i$ are integers.
\end{theorem}
\proof From Lemma \ref{le:7.1},  if $0\leq i\leq e-2, 1\leq j\leq2^{s-2}$, then each set $S_{ij}$ has two $q$-cylotomic coset. If $e-1\leq i\leq e$, then each set $S_{i0}$  has only one $q$-cylotomic coset. From Lemmas \ref{le:3.2}, \ref{le:7.1} and \ref{le:7.2}, then we have the number of $S_{ij}$ with $|S_{ij}|=1$ is 2, the number of $S_{ij}$ with $|S_{ij}|=2$ is $2^{s-1}-1$ and the number of $S_{ij}$ with $|S_{ij}|=2^{t-i+1}$ is $2^{s-2}$. Next we choose the appropriate defining set of cyclic code $C$ such that $ \ell=\sum_{i=0}^{t}n_i2^i.$ First we choose $n_0$ sets $S_{ij}$ with $|S_{ij}|=2$, where $0\leq n_0<2^{s-1}$. Let set $S_0$ be a union of one $q$-cylotomic coset of each $S_{ij}$. Second we choose $n_{i+1}$ sets $S_{ij}$ with  $|S_{ij}|=2^{t-i+1}$,  where $0\leq n_{i+1}\leq2^{s-2}$. Let  $S_{i+1}$ be a union of one $q$-cylotomic coset of each $S_{ij}$, where $0\leq i\leq t-1$. From the proof of Theorem \ref{th:7.3}, in addition to the first two steps, there are still ${(t+2)2^{s-2}-\sum_{i=0}^{t}n_i+1}$ sets $S_{ij}$ and let $S'_{ij}:=S_{ij}$. Finally let $S$ be a union of $S_i$ ($0\leq i\leq t$) and some of set $S'_{ij}$. Clearly, we have $\ell=|S\setminus(-S)|=\sum_{i=0}^{t}n_i2^i$, this is complete the proof. \qed

From the proof of Theorem \ref{th:7.4}, we know that we can obtain the enumeration of cyclic codes with $\ell={\rm dim(Hull}(C))\in N_1$. Next we only present the cyclic codes  with $\ell=1,2$.
\begin{theorem}\label{th:7.5}
	Let symbols be the same as before. Then the total number of cyclic codes with length $n$ and $\ell={\rm dim(Hull}(C))=1$ over $\mathbb{F}_q$ is equal to
	$$(2^{s}-2)2^{(t+2)2^{s-2}}.$$
\end{theorem}
\proof Let $S$ be the defining set of cyclic code $C$. We choose one $S_{i_0j_0}$ with $|S_{i_0j_0}|=2$, and one $q$-cyclotomic coset $C_0\in S_{i_0j_0}$. Suppose $S$ is a union of $C_0$ and some sets $S_{ij}$, where $(i,j)\neq(i_0,j_0)$. Clearly, we have $S\setminus(-S)=\{\frac{in}g\}$, for some $0\leq i\leq g-1$. By Lemma \ref{le:9.1}, The code $C$ has one-dimensional hull. From the proof of Theorems \ref{th:7.3} and \ref{th:7.4}, then the desired conclusion is obtained immediately. \qed
\begin{example}
	Let $(e,s)=(3,2)$, from Theorem \ref{th:7.5}, we know the total number of cyclic codes of length 8 and $\ell={\rm dim(Hull}(C))=1$ is 16. Let $S$ be defining set of cyclic code $C$ and $S=C_1\bigcup C_2\bigcup C_3\bigcup C_4$, from Theorem \ref{th:7.5}, we have that $C$  is cyclic code with one-dimensional hull and parameter $[8,2,6]$, which is optimal according to the Database.
\end{example}

\begin{theorem}\label{th:9.2}
	Let symbols be the same as before. Then the total number of cyclic codes with length $n$ and $\ell={\rm dim(Hull}(C))=2$ over $\mathbb{F}_q$ is equal to
	$$(2^{2s-2}-2^{s}+2)2^{(t+2)2^{s-2}}.$$
\end{theorem}
\proof It is similar to Theorem \ref{th:7.5}. \qed
\begin{example}
	Let $(e,s)=(4,2)$, from Theorem \ref{th:9.2}, we know the total number of cyclic codes of length 16 and $\ell={\rm dim(Hull}(C))=2$ is 32. Let $S$ be defining set of cyclic code $C$ and $S=C_6$, from Theorem \ref{th:9.2}, we have that $C$  is cyclic code with two-dimensional hull and parameter $[16,14,2]$, which is optimal according to the Database.
\end{example}
\subsection{The hull of cyclic code when $q=2^sf+3$}
\hspace{1.2 em}In this subsection, let $q=2^sf+3$. Let symbols be the same as Section 4, we have the following lemma immediately.

\begin{lemma}\label{le:8.1} For any $a\in T_i$, $C_a\in T_i$. We have following sentences are hold
	
	(1)  If $0\leq i\leq e-3$, $C_a \bigcap (-C_a)=\emptyset$;
	
	(2)  If $e-2\leq i\leq e$, $C_a\bigcap (-C_a)=C_a$.
	
	(3)  If $0\leq i\leq e$, $T_i\bigcap (-T_i)=T_i$.
\end{lemma}
\proof It is similar to Lemma \ref{le:7.1}. \qed

 From Lemmas \ref{le:5.1} and \ref{le:8.1}, if $0\leq i\leq e-3$, then each set $T_{i}$ has two $q$-cylotomic coset. If $e-2\leq i\leq e$, then each set $T_{i}$  has only one $q$-cylotomic coset. 

\begin{theorem}\label{th:8.2}
	Let symbols be the same as before. Then the total number of LCD cyclic codes with length $n$ over $\mathbb{F}_q$ is equal to
	$$2^{e+1}.$$
\end{theorem}
\proof It is similar to Theorem \ref{th:7.3}.\qed

\begin{theorem}\label{th:8.3}
	Let symbols be the same as before. Let $C$ be cyclic codes with length $n$ over $\mathbb{F}_q$,  $S$ denote the defining set of cyclic code $C$ and $\ell={\rm dim(Hull}($C$)).$ Then we have 
	$$ \ell\in N_2=\{~\sum_{i=1}^{e-1}n_i2^i~|~ 0\leq n_i\leq1~\},$$ where $n_i$ are integers.
\end{theorem}
\proof It is similar to Theorem \ref{th:7.4}.\qed

From Theorem \ref{th:8.3}, we know that we can obtain the enumeration of cyclic codes with $\ell={\rm dim(Hull}(C))\in N_2$ and following corollary.

\begin{corollary}\label{th:8.4}
There are no exist cyclic code of length $n$ over $\mathbb{F}_q$ with one-dimensional hull.
\end{corollary}

\begin{theorem}\label{th:9.3}
	Let symbols be the same as before. Then the total number of cyclic codes with length $n$ and $\ell={\rm dim(Hull}(C))=2$ over $\mathbb{F}_q$ is equal to
	$$2^{e+1}.$$
\end{theorem}
\proof It is similar to Theorem \ref{th:7.5}. \qed
\begin{example}
	Let $(e,s)=(4,3)$, from Theorem \ref{th:9.3}, we know the total number of cyclic codes of length 16 and $\ell={\rm dim(Hull}(C))=2$ is 32. Let $S$ be defining set of cyclic code $C$ and $S=C_1\bigcup C_2\bigcup C_4\bigcup C_5\bigcup C_8$, from Theorem \ref{th:9.3}, we have that $C$  is cyclic code with two-dimensional hull and parameter $[16,3,12]$, which is a almost MDS code.
\end{example}

\section{Conclusion}
\hspace{1.2 em}In this paper,  we studied the cyclic codes of length $n=2^e$ and dimension $k=2^{e-1}$. This $n$  is not satisfied with the condition of Ding's construction. We  gave two new types of generalized cyclotomy of order two, these new cyclotomys are different from Ding's. Furthermore, we  studied two classes of cyclic codes of length $n$ and dimension $k$. We got the enumeration of cyclic codes $C_{(n,\frac{n}{2},q)}$ over $\mathbb{F}_q$ with $q=2^sf+1$ and  $q=2^sf+3$,  respectively.  What's more, all of the codes from our construction are among the best cyclic codes. The 
equivalence of these cyclic codes from our construction will be considered in the future.  We construct and enumerate the cyclic codes of length $n$ having given dimensional hull. What's more, we obtain some optimal cyclic codes with small dimensional hull.


\begin{thebibliography}{199}
	\bibitem{ref8}    Betti E.  and Sala M., A new bound for the minimum distance of a cyclic code from its defining set, IEEE Trans. Inf. Theory, 52 (2006), 3700–3706.
		
	\bibitem{ref14} Baumert  L. D., Cyclic Difference Sets. Lecture Notes in Mathematics, no. 182. New York: Springer-Verlag, 1971.
		
	\bibitem{ref1}  Ding C. S. and Pless V., Cyclotomy and duadic codes of prime lengths, IEEE Trans. Inf. Theory, 45 (1999), 453-466.
	
	\bibitem{ref16} Ding C. S. and Helleseth T. , New generalized cyclotomy and its applications, Finite Fields Their Appl., 4 (1998) 140–166.
	
	\bibitem{ref2}  Ding C. S. and Helleseth T.,  Generalized cyclotomic codes of length $p_1^{e_1}\cdots p_t^{e_t}$, IEEE Trans. Inf. Theory, 45 (1999), 467-474.

	\bibitem{ref5}  Ding C. S., Cyclotomic constructions of cyclic codes with length being the product of two primes, IEEE Trans. Inf. Theory, 58 (2012), 2231-2236.
	
	\bibitem{ref3}  Ding C. S. and Xing C. P., Cyclotomic optical orthogonal codes of composite lengths, IEEE Trans. Inf. Theory, 52 (2004), 263-268.
	
    \bibitem{ref4}  Ding C. S. and  Niederreiter H., Cyclotomic linear codes of order 3, IEEE Trans. Inf. Theory, 53 (2007), 2274-2277.
    
	\bibitem{ref13}  Gauss C. F., Disquisitiones Arithmeticae. Leipzig, Germany, 1801; English translation: New Haven, CT: Yale Univ. Press, 1966; reprinted: Berlin, Heidelberg, and New York: Springer-Verlag, 1986.
	
	\bibitem{ref25} Ghinelli D.,  Key J. D. and  McDonough T. P., Hulls of codes from incidence
	matrices of connected regular graphs, Des. Codes Cryptogr., 70 (2014), 35–54.
	
    \bibitem{ref9}  Huffman W. C. and  Pless V., Fundamentals of Error-Correcting Codes. Cambridge, U.K.: Cambridge Univ. Press, 2003.
    
   	\bibitem{ref10} Jia Y., Ling S. and Xing C. P., On self-dual cyclic codes over finite fields, IEEE Trans. Inf. Theory, 57 (2011), 2243–2251.
   	
   	\bibitem{ref17} Ko C. and Sun Q., Number theory lecture, Higher education press, China, 2012.
	
	
	\bibitem{ref7} Kewat P. K. and Kumari P., Cyclic codes from the second class two-prime Whiteman’s generalized cyclotomic sequence with order 6, Cryptogr. Commun., 9 (2017), 475-499.

	\bibitem{ref24}  Li C. J. and Zeng P., Constructions of linear codes with one-dimensional hull, IEEE Trans. Inf. Theory, DOI 10.1109/TIT.2018.2863693.
	
	\bibitem{ref18} Leon J., Computing automorphism groups of error-correcting codes,
	IEEE Trans. Inf. Theory,  3 (1982), 496–511.
	
	\bibitem{ref19} Leon J., Permutation group algorithms based on partition, I: theory and algorithms, J. Symbolic Comput., 12 (1991), 533–583.
	
	\bibitem{ref11}  Moisio M., Exponential sums, Gauss sums and cyclic codes, Ph.D. dissertation, Dept. Math. Sci., Univ. Oulu, Oulu, Finland, 1998.
	
	\bibitem{ref15}  Storer  T., Cyclotomy and Difference Sets. Chicago, IL: Markham, 1967.
	
	\bibitem{ref20} Sendrier N., On the dimension of the hull, SIAM J. Discrete Math., 10 (1997), 282–293.
	\bibitem{ref21} Sendrier  N. and  Skersys G., On the computation of the automorphism
	group of a linear code, in: Proceedings of IEEE ISIT’2001, Washington, DC, p. 13, 2001.
	
	\bibitem{ref22}  Sendrier N., Finding the permutation between equivalent codes: the support splitting algorithm, IEEE Trans. Inf. Theory, 46 (2000), 1193–1203.
	
	\bibitem{ref23}Skersys G., The average dimension of the hull of cyclic codes, Discrete
	Appl. Math., 128 (2003),  275–292.
	
	\bibitem{ref26} Sangwisut E.,  Jitman S.,  Ling S. and  Udomkavanich P., Hulls of cyclic
	and negacyclic codes over finite fields, Finite Fields Appl., 33 (2015), 232–257.
	
	\bibitem{ref12} van Lint  J. H. and  Wilson R. M., On the minimum distance of cyclic codes, IEEE Trans. Inf. Theory, 32 (1986), 23–40.
	
	\bibitem{ref6}  Xiong M. S., On cyclic codes of composite length and the minimum distance, IEEE Trans. Inf. Theory, DOI 10.1109/TIT.2018.2829763.
	


\end{thebibliography}
\end{document}